\PassOptionsToPackage{usenames,dvipsnames}{xcolor}
\PassOptionsToPackage{colorlinks=true,citecolor=magenta,linkcolor=magenta,urlcolor=magenta}{hyperref}

\documentclass[
  sigconf,
  natbib=false,   
  usenames,
  dvipsnames,
  nonacm,
  10pt
]{acmart}

\makeatletter                   
\def\mdseries@tt{m}             
\makeatother                    

\usepackage[frozencache,cachedir=.,draft=true]{minted}     

\usepackage[medium]{titlesec}
\titlespacing*{\section}{0pt}{3pt}{3pt}
\titlespacing*{\subsection}{0pt}{3pt}{3pt}

\usepackage[aboveskip=2pt,belowskip=0pt]{caption}

\usepackage{balance}

\usepackage{algorithm}
\usepackage[noend]{algpseudocode}   

\makeatletter

\algrenewcommand\ALG@beginalgorithmic{\small}  
\algrenewcommand\algorithmiccomment[1]{\textcolor{OliveGreen}{// {\itshape #1}}}

\algnewcommand{\algorithmicand}{\textbf{ and }}
\algnewcommand{\algorithmicor}{\textbf{ or }}
\algnewcommand{\OR}{\algorithmicor}
\algnewcommand{\AND}{\algorithmicand}

\algblock{ParFor}{EndParFor}
\algnewcommand\algorithmicparfor{\textbf{parallel for}}
\algnewcommand\algorithmicpardo{\textbf{do}}
\algnewcommand\algorithmicendparfor{\textbf{end\ parallel for}}
\algrenewtext{ParFor}[1]{\algorithmicparfor\ #1\ \algorithmicpardo}
\algrenewtext{EndParFor}{\algorithmicendparfor}

\makeatother

\usepackage{textcomp}
\usepackage{graphicx}
\usepackage{epstopdf}
\usepackage{listings}
\usepackage{xspace}

\usepackage{xurl}
\usepackage[colorlinks=true,citecolor=magenta,linkcolor=magenta,urlcolor=magenta]{hyperref}

\usepackage{booktabs}
\usepackage{multirow}

\usepackage{minted}

\usepackage{outlines}

\usepackage{soul}

\usepackage{todonotes}

\usepackage{hyphenat}

\usepackage{amsmath}

\usepackage{enumitem}
\setlist[itemize]{leftmargin=*}
\setlist[enumerate]{leftmargin=*}
\setlist[description]{leftmargin=3mm}

\usepackage[nodisplayskipstretch]{setspace}

\usepackage{tabularx}
\newenvironment{conditions*}
  {\par\vspace{\abovedisplayskip}\noindent
   \tabularx{\columnwidth}{>{$}l<{$} @{${}={}$} >{\raggedright\arraybackslash}X}}
  {\endtabularx\par\vspace{\belowdisplayskip}}

\usepackage[firstinits=true,abbreviate=true,dateabbrev=true,url=true,maxbibnames=9,language=american,backend=biber,sorting=none, language=american]{biblatex}
\usepackage{enumitem}
\setlist[itemize]{leftmargin=*}
\setlist[enumerate]{leftmargin=*}
\setlist[description]{leftmargin=3mm}

\usepackage{subcaption}
\newcommand{\ccc}[1]{}  

\newcommand{\goner}[1]{} 

\newcommand{\sysname}{\textsc{Kairos}\xspace}  
\newcommand{\DS}{\textsc{TGER}\xspace}  

\newcommand{\mypara}[1]{\vspace{2mm}\noindent\textbf{#1}}

\renewcommand{\footnotesize}{\scriptsize}

\newmintedfile[sqlcode]{sql}{
  fontsize=\scriptsize,
}

\setcopyright{rightsretained}

\acmDOI{10.475/123_4}

\acmISBN{123-4567-24-567/08/06}

\begin{document}


\title{Kairos: Efficient Temporal Graph Analytics\\ on a Single Machine}

\author{Joana M. F. da Trindade}
\affiliation{\institution{MIT CSAIL}}
\email{jmf@csail.mit.edu}
\author{Julian Shun}
\affiliation{\institution{MIT CSAIL}}
\email{jshun@csail.mit.edu}
\author{Samuel Madden}
\affiliation{\institution{MIT CSAIL}}
\email{madden@csail.mit.edu}
\author{Nesime Tatbul}
\affiliation{\institution{Intel Labs / MIT CSAIL}}
\email{tatbul@csail.mit.edu}

\renewcommand{\shortauthors}{J. M. F. da Trindade et al.}


\begin{abstract}
  Many important societal problems are naturally modeled as algorithms over temporal graphs. To date, however, most graph processing systems remain inefficient as they rely on distributed processing even for graphs that fit well within a commodity server's available storage. In this paper, we introduce \sysname, a temporal graph analytics system that provides application developers a framework for efficiently implementing and executing algorithms over temporal graphs on a single machine.
Specifically, \sysname relies on fork-join parallelism and a highly optimized parallel data structure as core primitives to maximize performance of graph processing tasks needed for temporal graph analytics.
Furthermore, we introduce the notion of \emph{selective indexing} and show how it can be used with an efficient index to speedup temporal queries.  Our experiments on a 24-core server show that our algorithms obtain good parallel speedups, and are significantly faster than equivalent algorithms in existing temporal graph processing systems: up to 60x against a shared-memory approach, and several orders of magnitude when compared with distributed processing of graphs that fit within a single server.
\end{abstract}

\maketitle



\section{Introduction}
\label{s:intro}

The growing demand for temporal graph applications has given rise to new challenges in temporal graph analytics. As an increasing number of real-world systems and processes can be modeled as temporal graphs, the need for effective analysis tools and techniques has become more pressing. These applications range from social networks and communication systems to transportation networks and biological systems, where understanding the temporal dynamics is crucial for uncovering meaningful insights and patterns~\cite{TemporalNetworks, Epi1, Motifs2, TGLBenchmark}.

Temporal graphs offer a unique perspective as they can capture the dynamics of interactions and relationships over time, which non-temporal graphs are unable to provide. Furthermore, temporal graphs lend themselves to the exploration of time-ordered events and the impact of such sequences in the network, creating an opportunity for more nuanced insights. For instance, being able to track the chronology of friendship formations on a social network or communication events can enhance our understanding of behavioral patterns. Similarly, the ability to observe the evolution of transportation routes or the progression of a biological system over time can provide critical data for predictive modeling and decision-making.

\begin{figure}[t!]
 \centering\includegraphics[width=0.35\textwidth]{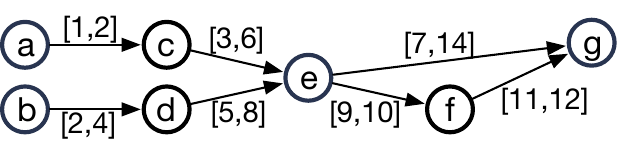}
 \caption{An example temporal graph representing vertices $\{a, b, c, d, e, f, g\}$.
 Each edge is associated with a time interval (start, end) denoting its validity.}
 \label{fig:temporal_graph}
\end{figure}

However, the temporal dynamics inherent in these graphs also brings about new challenges. Existing graph frameworks and query systems frequently encounter difficulties when handling graph processing tasks required in temporal graph analytics applications. The underlying reasons for these challenges are twofold. First, many of these systems were primarily designed for traditional graph processing, and as a result, they are not well-equipped to handle the unique characteristics and requirements of temporal graphs. This shortcoming leads to suboptimal performance and a limited ability to fully leverage the available temporal information in the data. Second, some systems~\cite{ICM, Tink, GRADOOP} that specifically target temporal graph processing rely on Pregel-like distributed computation models. These models can be highly inefficient due to the message passing overhead across servers in a cluster, especially when the input graph fits comfortably within the memory resources of a single commodity machine. This limitation leads to suboptimal performance and an inability to fully exploit the temporal information available in the data. Finally, existing shared-memory temporal graph processing systems~\cite{TeGraph} do not have these limitations, but still rely on expensive pre-processing of the graph that increases the original size of the dataset -- a step which is not necessary for correctness of the target algorithms.

Temporal graph applications typically require querying small time slices of data. While existing systems can represent time as an attribute of nodes and edges, filtering by time necessitates either an expensive scan or the use of a range index, resulting in poor performance. Furthermore, traditional graph processing frameworks lack support for temporal algorithms, leading to complex and costly implementations as they require additional programming effort.

Temporal graphs also exhibit unique properties that are not necessarily considered in existing systems. For example, they  tend to be large due to the increased number of attributes and edges, resulting from the preservation of interactions between vertices over time.
Additionally, temporal graphs exhibit skewed data distributions, both in terms of degree distributions and their evolution over time~\cite{leskovec2005graphs, TemporalNetworks}. To address these challenges, temporal graph analytics systems must provide interactive response times for various use cases, such as operational decisions, contact tracing, and routing.

\mypara{Our Approach and Contributions.} In response to the challenges described above, we have developed \sysname, an in-memory temporal graph analytics system that leverages a highly-optimized parallel data structure for to efficiently execute temporal algorithms over temporal graphs on multi-core machines. \sysname is designed to address the unique properties and requirements of temporal graphs, providing an efficient solution for processing large-scale temporal graphs. Specifically, it introduces programming primitives and APIs that organize temporal graphs by time, as well as a novel Temporal Graph Index designed for efficiently looking up data in specific time ranges. This approach enables efficient temporal graph algorithm implementations, treating temporal edge information as first-class citizens in the model. The system also incorporates selective indexing, a query optimization technique that relies on a novel cost model to speed up query execution by choosing the best access method for neighbors of a given vertex at runtime.

We have implemented a number of parallel temporal graph algorithms for various application classes, including single-source shortest paths (earliest arrival, latest departure, fastest, and shortest duration), connectivity (temporal connected components), and centrality (temporal betweenness centrality). Our system provides significant performance improvements compared to existing state-of-the-art temporal graph processing systems. We make the following contributions:

\begin{enumerate}[label=\Roman*.]
\item We present \DS, a novel ``time-first'' data structure that acts as an index to enable efficient processing of temporal graph queries and algorithms. 

\item We introduce {\em selective indexing}, a technique that speeds up query execution by choosing the best access method for retrieving neighbors of a given vertex at runtime.

\item We present efficient shared-memory parallel algorithm implementations for various temporal graph application classes, offering significant performance improvements compared to existing state-of-the-art temporal graph processing systems.

\item Our results show substantial speedup compared to existing systems and provide insights into the performance characteristics and guarantees of our system.
\end{enumerate}


\section{Background}
\label{s:background}

\begin{table*}[t!]
\small
\centering
\setlength\tabcolsep{1pt} 

\resizebox{0.85\textwidth}{!}{%

\begin{tabular}{p{3.8cm}p{13.5cm}}
\hline
Category & Specific instantiation\\
\hline
\multicolumn{2}{c}{\textbf{Temporal Graph Algorithms}} \\\hline
Temporal Minimal Paths
& [Provenance] Tracking the origin and flow of information in different systems \cite{Provenance1, Provenance2} \\
& [Indoor Routing] Temporal shortest paths that consider different obstacles for correct navigation~\cite{Indoor1, Indoor2} \\
& [Transportation] Route planning that considers real-time traffic conditions in transportation networks \cite{Minimal1} \\
& [Epidemiology] Identifying infection transmission paths in contact tracing \cite{Epi1} \\\hline
Temporal Connectivity
& [Social Networks] Analyzing community evolution and detecting temporal clusters \cite{Clusters1, Clusters2, Clusters3, Clusters4} \\\hline
Temporal Centrality
& [Epidemiology] Identifying critical nodes in the spread of infections \cite{Centrality1} \\\hline

\multicolumn{2}{c}{\textbf{Temporal Graph Queries}} \\\hline
Time-constrained reachability & [Social Networks] Analyzing influence propagation and information cascades \cite{InfoPropagation1, InfoPropagation2, Reachability1} \\
Temporal subgraph matching & [Bioinformatics] Detecting conserved patterns in dynamic biological networks \cite{Motifs1, Motifs2} \\\hline

\end{tabular}

} 

\caption{A Categorization of tasks commonly performed in temporal graph analytics.}
\label{tab:example_apps}
\end{table*}

\subsection{Temporal Graph Data Model}
\label{ss:dm}

A temporal graph is represented by the tuple $G = (V, E, T, \tau)$:
\begin{itemize}

\item $V$ denotes a set of vertices.
\item $E$ denotes a set of edges.
\item $T = [0, 1, ..., t_{max}] \in \mathbb{N}$ represents a discrete time domain.
\item $\tau : V \times V \times T \times T \to \{ False,True \}$ is a function that determines for each pair of vertices $u, v \in V$, and each pair of timestamps $t_{start}, t_{end} \in T$ where $t_{start} \leq t_{end}$, whether $(u, v) \in E$, i.e., whether the edge $(u, v)$ exists during the discrete time period from $t_{start}$ to $t_{end}$.

\end{itemize}
  In other words, each edge in a temporal graph is associated with a discrete time interval indicating its validity. For instance, in interaction networks, this time interval indicates the period during which two vertices have interacted.

A weighted temporal graph is represented by the tuple $G = (V, E, T, \tau, w)$, where $w$ is a function that maps a temporal edge (as defined above) to a real value (its weight). The number of vertices in a temporal graph is $n_{v} = |V|$, and the number of edges is $n_{e} = |E|$. Vertices are assumed to be labeled from $0$ to $n_{v} - 1$. For undirected temporal graphs, we use $deg(v)$ to denote the number of edges incident to a vertex $v \in V$. In directed temporal graphs, vertices contain both incoming and outgoing edges. We use $\textit{deg}^{+}(v)$ to denote the number of outgoing edges, and $\textit{deg}^{-}(v)$ to denote the number of incoming edges that a vertex $v$ has.


\subsection{Allen's Interval Algebra}
\label{ss:interval_algebra}

We draw inspiration from Allen's Intervala Algebra~\cite{interval_algebra} to specify the relationships that subsequent edges in the same path must have. The following subset of this algebra defines the validity of temporal paths:

\begin{itemize}
\item \textbf{Succeeds:} For two time intervals $A$ and $B$, $B$ \textit{succeeds} $A$ if and only if the end time of $A$ is smaller than or equal to the start time of $B$, i.e., $\textit{end}(A) \leq \textit{start}(B)$.
\item \textbf{Strictly succeeds:} $B$ \textit{strictly succeeds} $A$ if and only if the end time of $A$ is strictly smaller than the start time of B, i.e., $\textit{end}(A) < \textit{start}(B)$.
\item \textbf{Overlaps:} $B$ \textit{overlaps} $A$ if and only if the start time of $A$ is less than the start time of $B$, and the end time of $A$ is less than the end time of $B$, i.e., $\textit{start}(A) \le \textit{end}(B)$ and $\textit{end}(A) \le \textit{start}(B)$.
\end{itemize}

We refer to this subset as {\em ordering predicates} in \sysname, and describe its application in~\autoref{ss:interface}.


\subsection{Temporal Graph Analytics Tasks}
\label{ss:temporal_algos}

In Section \ref{s:intro}, we outlined several use cases for temporal graph analytics. Here, we provide a survey of applications of temporal graph analytics algorithms and queries from the literature, listed in \autoref{tab:example_apps}. For each algorithm and query, we present an example application that relies on it as a core primitive for analysis.  We focus on a core set of algorithms that address the most common use cases in temporal graph analytics and can serve as primitives for building more advanced analysis tasks. The primary categories for these algorithms are temporal paths, temporal connectivity, and temporal centrality.

\mypara{Temporal Paths:} A temporal path in graph $G$ is a path where every subsequent edge in the path must satisfy certain temporal constraints. Examples of minimal temporal paths include earliest arrival, latest departure, fastest path, and shortest path~\cite{TemporalPaths, TemporalPathsJournal}.

\mypara{Temporal Connectivity:} Temporal connectivity deals with the temporal version of connected components. This involves identifying sets of vertices that are connected through time.

\mypara{Temporal Centrality:} Temporal centrality measures the importance of a vertex in a temporal graph. An example of interest is temporal betweenness centrality, which quantifies how frequently a vertex appears on temporal shortest paths between that vertex and other vertices in the graph.


\subsection{The Compressed Sparse Row (CSR) Format}
\label{ss:csr}


In the context of parallel graph processing, a common data structure used to store graph data is the Compressed Sparse Row (CSR) format. This format is largely preferred over other competing data structures as it allows for efficient storage and manipulation of sparse graphs, in which the majority of potential edges are absent~\cite{packed_CSR}.

The CSR representation of a graph is characterized by the use of three arrays: an adjacency array, an offset array, and a vertex array. The adjacency array serves as a storage mechanism for destination vertices of all edges in a graph. The edges are placed in a contiguous block of memory, where they are sorted by the source vertex for outgoing edges, and sorted by the destination vertex for incoming edges. This efficient organization supports the fast retrieval of adjacent vertices in graph traversal operations, as all vertices in the same 1-hop neighborhood are located contiguously in memory. The offset array then stores the indices of outgoing / incoming edges into the adjacency array. These indices mark the starting point of the adjacency list for each vertex, enabling quick access to the set of edges associated with any given vertex. Lastly, if there is no metadata associated to vertices, then the vertex id is implicit (i.e., offset[$i$] contains the offset for vertex $i$), and no separate vertex ids array is needed, which is the case for \sysname.


\subsection{Shared-memory Graph Processing: Ligra}
\label{ss:ligra_prim}

Ligra is a lightweight graph processing framework designed for shared-memory parallel systems~\cite{ligra}. It enables efficient parallel graph processing by employing a simple and flexible programming model, making it easy for developers to write high-performance algorithms for large-scale graphs.

The programming model of Ligra is centered around two primary operations: \textit{EdgeMap} and \textit{VertexMap}. These operations enable parallel traversal of the graph and are responsible for most of the computation in a Ligra-based algorithm.

\textbf{EdgeMap} is a higher-order function that takes as input a graph $G$, a subset of vertices $V'$, and an edge function $f$. It applies the edge function $f$ to all edges $(u,v)$ in the graph, where $u \in V'$ and $v \in V$. The edge function $f$ is responsible for implementing the logic of the specific graph algorithm and can perform various operations, such as updating vertex properties or computing edge weights. EdgeMap efficiently handles parallelism by processing edges in parallel, allowing for scalable performance on shared-memory systems.

\textbf{VertexMap} is another higher-order function that takes as input a graph $G$, a subset of vertices $V'$, and a vertex function $g$. It applies the vertex function $g$ to all vertices in $V'$. Similar to the edge function, the vertex function is responsible for implementing the algorithm-specific logic and can perform operations such as updating vertex properties or aggregating information from neighboring vertices. VertexMap also processes vertices in parallel, ensuring scalable performance.

By using these two core operations, Ligra enables developers to write graph algorithms that can efficiently exploit the parallelism offered by shared-memory systems. In the context of \sysname, we build upon the Ligra programming model and extend it to support temporal graph operations, as we will discuss in the following sections.


\subsection{Problem Formulation}
\label{ss:problem_formulation}

The primary challenge in designing a temporal graph analytics system is to efficiently process and analyze temporal graphs while taking into account the unique characteristics of such graphs, such as time-varying edges and vertices. The problem can be stated as follows: given a temporal graph $G = (V, E, T, \tau)$, our goal is to efficiently support the execution of temporal graph analytics tasks, including temporal paths, temporal connectivity, and temporal centrality.

The example queries and algorithms discussed above require different strategies for selecting, at runtime, the appropriate set of vertices and temporal edges to be explored in the frontier being traversed by the graph processing engine. The efficiency of these strategies is crucial for the performance of temporal graph analytics algorithms. In this paper, we aim to address this challenge by introducing selective indexing and other design decisions that help significantly improve the performance of executing different tasks over an input temporal graph. To address this problem, we must fulfill the following desiderata:

\begin{enumerate}
\item Efficient storage and indexing of temporal graph data, especially as it pertains to temporal edges.
\item Design and implementation of efficient parallel algorithms for temporal graph analytics tasks, such as temporal paths, temporal connectivity, and temporal centrality.
\item Effective runtime selection of the edges to be explored in the frontier to minimize the time and space complexity of the operations involved in the analytics tasks.
\item Provide a flexible and easy-to-use programming model to support the implementation of various temporal graph analytics tasks and queries.
\end{enumerate}

By addressing these aspects, our goal with \sysname is to provide an efficient temporal graph analytics system that can handle large-scale temporal graphs and deliver appropriate programming primitives for writing temporal graph applications.

\section{Architectural Overview and Key Ideas}
\label{s:overview}

\sysname provides an API that is compatible with that of a state-of-the-art high-performance graph processing system (\cite{ligra}), and extends it to the temporal setting. The API provides methods to load graphs, perform operations such as traversal, filtering, and allows computations to be expressed in terms of an input temporal query. 

\begin{figure}[!t]
  \begin{center}
  \includegraphics[width=0.8\columnwidth]{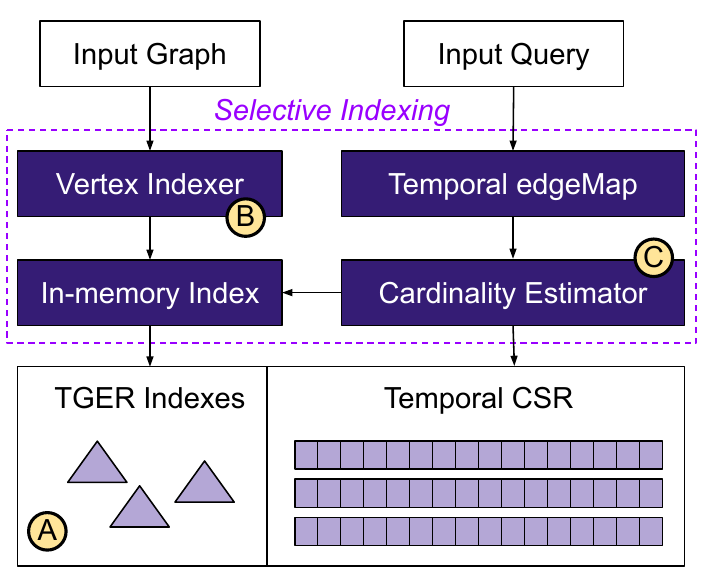}
  \caption{Architectural overview and key ideas.}
  \label{fig:arch}
  \end{center}
\end{figure}

In the subsequent sections, we will delve deeper into the key ideas behind our approach, as outlined in \sysname's architecture diagram (\autoref{fig:arch}). We introduce \sysname's core data structures, the need to choose between different access methods, as well as the types of information needed to make this decision. First, we describe a high-specialized parallel data structure for indexing temporal edges (\autoref{ss:overview-specialized_indexes}). Second, we introduce novel approach to {\em selectively} decide which subset of vertices are benefitial to index, as well as how to access each vertex at runtime (\autoref{ss:overview-vertex_indexing}) and associated cost model.

\subsection{Temporal Graph Edge Registry (Key Idea A)}
\label{ss:overview-specialized_indexes}

\begin{figure*}[htb!]
  \begin{center}
  \includegraphics[width=0.9\textwidth]{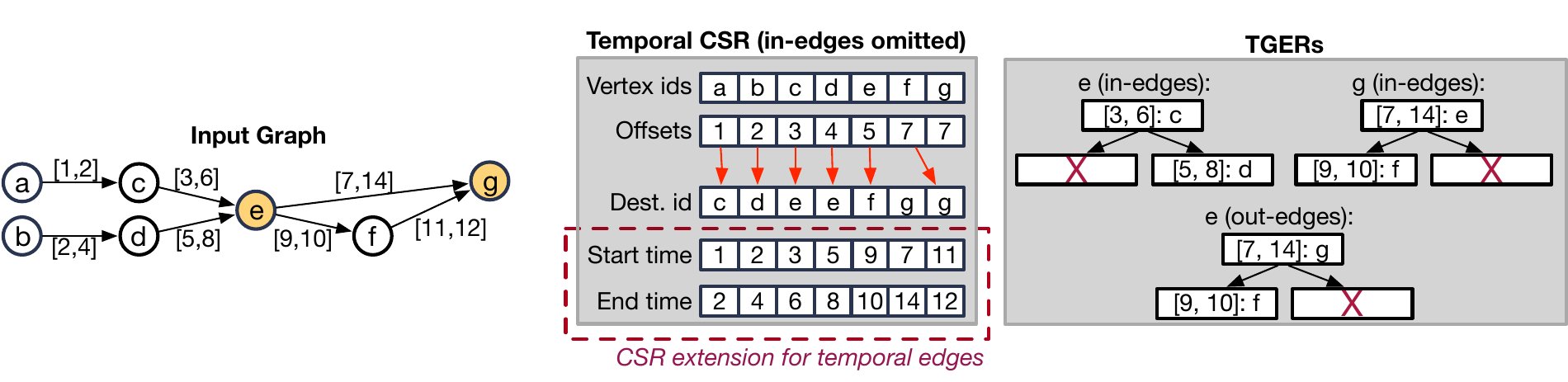}
  \caption{Data layout for example temporal graph in \autoref{fig:temporal_graph}. Temporal CSR is available for all vertices in the graph (incoming edges omitted for sake of clarity). In this example, TGER is used to index only the vertices with out or in-degree of at least two edges. In practice, \sysname only indexes vertices with much larger degrees.}
  \label{fig:temporal_csr_and_TGER}
  \end{center}
\end{figure*}

Temporal graphs introduce an additional layer of complexity in graph computation due to the addition of time as a variable. To process queries and algorithtms over temporal graphs, one could consider a naive approach based on CSR (\autoref{ss:csr}). With this approach, the entire list of edges associated to a given vertex is stored in the corresponding offset of the CSR, and then a parallel filter is applied to retrieve neighbors satisfying an input temporal predicate. This can become expensive, however, particularly in graphs with high-degree vertices, or when the output of this filtering is much smaller than the degree of the vertex. This can decrease performance, as a large number of edges that are not relevant to the specific input temporal predicate might need to be processed.

Therefore, the efficient processing of queries and temporal algorithms over temporal graphs demands specialized data structures for storing and retrieving vertex neighbors. To this end, we have designed and implemented \DS (Temporal Graph Edges Registry), a parallel data structure based on priority search trees~\cite{CompGeometryBook} for effectively processing queries and temporal algorithms over temporal graphs. \DS lets \sysname treat temporal edge information as first-class citizens in the data model, enabling efficient temporal graph algorithm implementations. Moreover, \DS is storage-efficient (i.e., $O(m)$ space, where $m$ is the number of temporal edges stored in it) and can answer interval containment queries efficiently ($O(\log m + k)$ work, where $k$ is the number of results~\cite{CompGeometryBook}). Finally, \DS makes efficient use of computational resources, employing fork-join parallelism in its implementation of both construction and query operations. As we show in our experiments, \DS can provide a substantial advantage over the naive CSR-based approach.


\subsection{Selective Indexing (Key Ideas B and C)}
\label{ss:overview-vertex_indexing}

The efficient retrieval of vertices and edges of interest is crucial for the performance of algorithms and queries in large graphs. This is particularly true for large-scale real-world graphs, where data is often skewed, and a small number of vertices can account for most edges. The skewness is intensified in real-world temporal graphs, where data can also be unevenly distributed over time (e.g., seasonal data patterns, or growth in popularity in the case of social networks). Furthermore, the skew present in  real-world temporal graphs is a well studied phenomena, and which can be attributed to inter-contact time distributions, burstiness, and even circadian or otherwise weekly rhythms that are inherent in human activity~\cite{TemporalNetworks}. For this reason, different approaches have been proposed to generate temporal graphs that are closer to real-world skewed distributions~\cite{TACO_VLDB2022, TREND_WWW2022}, and more recently data imputation~\cite{AdaptingToSkew} that can handle skew.

To address this challenge, we introduce a new class of problems in query optimization for graph queries, as well as a unique technique for {\em selectively} indexing different parts of a graph relevant to query answering. As part of our approach, we present a novel cost model and related algorithms for determining the most suitable access method (e.g., index-based vs.\ scan-based) for graph traversal. Our algorithms take into account the characteristics of the underlying data (e.g., data skew) and workload (e.g., selectivity of temporal predicates present in the queries) to choose the most efficient access method (index vs.\ scan) for each vertex at runtime. ~\autoref{s:indexing} (Selective Indexing) provides a description of our technique.

\mypara{Vertex Indexer.} This component is responsible for deciding which vertices warrant a \DS index. First, a vertex size is defined in terms of the size of their out-degree or in-degree neighborhood (i.e., how many outgoing or incoming edges it has). Based on a predefined vertex size threshold (heuristically obtained via experimental analysis), the indexer builds a \DS only for those vertices whose size exceeds this threshold. To quickly identify which vertices have a \DS, the Vertex Indexer maintains an in-memory sparse associative array where each entry maps a vertex id to the in-memory location of its corresponding \DS. At runtime, \sysname employs a cost model (~\autoref{s:indexing}) to evaluate whether it should access the corresponding \DS for that vertex, or opt for a linear scan using Temporal CSR (T-CSR, ~\autoref{ss:temporal_csr}).


\mypara{Cardinality Estimator.} We introduce a cost model for deciding at runtime which access method to use for retrieving a vertex's neighborhood. As part of this cost model, a cardinality estimator plays a key role in determining whether a query is selective enough to warrant index-based retrieval of outgoing/incoming edges via \DS. We further describe it and its associated algorithms in \autoref{ss:cost_model} and \autoref{ss:estimation_algo}.

\section{\sysname Framework}
\label{s:impl}

This section describes the interface and implementation of \sysname framework, with focus on how it extends Ligra~\cite{ligra} to the temporal setting. \autoref{tab:api} summarizes its interface.

\subsection{Interface}
\label{ss:interface}

\sysname contains two key data structures: \textbf{{\em VertexSet}} and \textbf{{\em TemporalEdgeSet}}, which are used respectively to represent subset of vertices and subset of temporal edges. \textbf{{\em OrderingPredicate}} takes as input a pair of temporal edges, an \textbf{{\em OrderingPredicateType}}, and evaluates whether the given temporal edges conform to that predicate (\autoref{ss:temporal_index}). Using these as input, \textbf{{\em TemporalGraph}} constructs the temporal graph. \textbf{{\em VertexMap}} is equivalent to Ligra's (\autoref{ss:ligra_prim}), with the only difference that it takes a \textbf{{\em TemporalGraph}} as input. Conversely, \textbf{{\em TemporalEdgeMap}} extends Ligra's \textbf{{\em EdgeMap}} to the temporal setting, and is described in more details in \autoref{ss:temporal_edgemap}).

\begin{figure}[htb!]
\centering\includegraphics[width=0.45\textwidth]{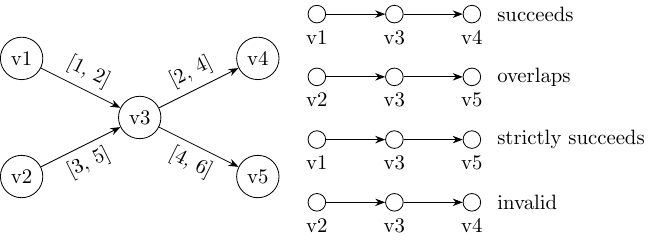}
\caption{Ordering predicates for each of the four possible two-hop directed paths on an example graph.}
\label{fig:ordering_predicates}
\end{figure}

\begin{table*}[t]
 \begin{center}

\resizebox{\textwidth}{!}{%

     \begin{tabular}{p{11cm}p{12cm}}
     \toprule
     \textbf{Interface} & \textbf{Description}\\

     \midrule
     VertexSet & Represents a subset of vertices $V' \subseteq V$.\\
     TemporalEdgeSet & Represents a subset of temporaledges $E' \subseteq E$.\\

     \midrule
     \textproc{OrderingPredicate}($A$: {\em temporaledge}, $B$: {\em temporaledge}, $T$: OrderingPredicateType): {\em bool}
     & Evaluates the order between two temporal edges based on one of the three ordering predicate types: Suceeds, StrictlySucceeds, or Overlaps (\autoref{ss:interface}). \\

     \midrule
     \textproc{TemporalGraph}($V$: VertexSet, $E$: TemporalEdgeSet, $P$: OrderingPredicate)
     & Constructs a temporal graph using given input vertices, temporaledges, and ordering predicate. \\

     \midrule
     \textproc{VertexMap}($U$: VertexSet, $G$: TemporalGraph, $F$: {\em vertex} $\to$ {\em bool}): VertexSet
     & Applies $F(u)$ for each $u \in U$; returns a VertexSet $\{u \in\ U \mid F(u) = \textit{true}\}$. \\

     \midrule
     \textproc{TemporalEdgeMap}($U$: TemporalEdgeSet, $G$: TemporalGraph, $F$: {\em temporaledge} $\to$ {\em bool}): TemporalEdgeSet
     & Applies $F(u)$ for each $u \in U$; returns a TemporalEdgeSet $\{u \in\ U \mid F(u) = \textit{true}\}$. \\

     \bottomrule
    \end{tabular}

} 

    \caption{Core primitives from \sysname framework programming model interface.}
    \label{tab:api}
 \end{center}  
\end{table*}


\subsection{T-CSR: Temporal CSR}
\label{ss:temporal_csr}

The Temporal CSR (T-CSR) data structure is an extension of the traditional CSR representation (\autoref{ss:csr}) designed to accommodate temporal graph data. It retains the core concepts of CSR, while incorporating additional arrays to store the start and end times associated with each temporal edge, as shown in \autoref{fig:temporal_csr_and_TGER}. Specifically, it extends the standard CSR representation by adding two more arrays, the start time array, and the end time array. As their name suggests, these arrays store the start and end times of each temporal edge in the graph, respectively. They are organized in the same order as the adjacency array, such that the start and end times of the temporal edge at position $i$ in the adjacency array can be found at position $i$ in the start / end time arrays.

With this extended representation, the T-CSR can store temporal graphs and support various temporal graph processing tasks. The additional start time and end time arrays enable the system to take into account the temporal information of the graph when executing queries and algorithms when used in conjunction with \textbf{{\em TemporalEdgeMap}} (\autoref{ss:temporal_edgemap}). The T-CSR representation retains the advantages of the traditional CSR, such as cache efficiency and fast access to adjacency lists. Furthermore, it can be easily incorporated into existing CSR-based graph processing systems with minimal modifications, such as is the case for our extension of Ligra to the temporal setting. Overall, the Temporal CSR offers an effective and efficient solution for storing and processing temporal graph data. Because of that, it is the default representation used to store the neighbors of most vertices in \sysname. A similar representation is also used in the Temporal GNN literature~\cite{TGL} for sampling temporal edges.


\subsection{\DS: Temporal Graph Edges Registry}
\label{ss:temporal_index}

While T-CSR offers valuable improvements over traditional CSR for temporal graph data handling, its design is still geared towards adjacency-oriented queries and operations. For example, T-CSR does an excellent job when dealing with questions of ``who is connected to whom''. However, it is less efficient when it comes to handling range-based temporal queries such as ``who was connected to whom within a specific time window''. This is because in T-CSR, the temporal information is appended to the existing structure, which is space-optimized for traditional graphs (i.e., where edges do not have additional temporal information). As a result, each temporal query needs to scan over the entire adjacency list of a vertex and filter out the relevant edges based on their timestamps. While this operation can be performed in parallel, it can still be inefficient for large graphs, or for queries with small time windows (i.e., queries that are highly selective) relative to the total neighbors a vertex has.

\begin{figure}[t!]
  \begin{center}
  \includegraphics[width=1\columnwidth]{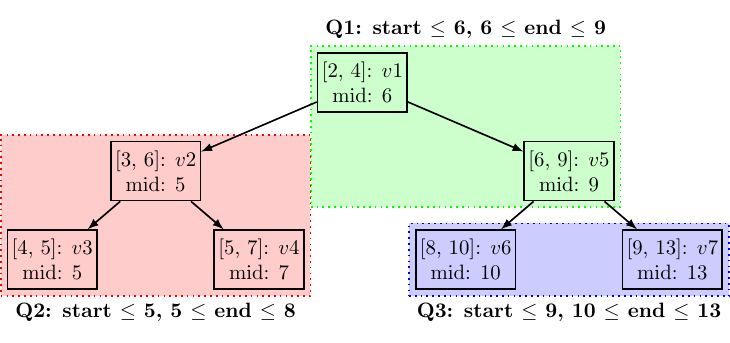}
  \caption{Queries in \DS are 3-sided. By default, an upper bound ({\em min-heap mode}) is specified for start time, and a lower plus upper bound is specified for end time.}
  \label{fig:index-queries}
  \end{center}
\end{figure}

To address this challenge, we propose \DS, an index data structure which accommodates temporal information (in the form of time intervals associated with edges) as a first-class citizen. TGER is a dual index, meaning it indexes both the start and end time attributes. Specifically, it combines a priority queue (by defaullt over the start time attribute) and a Binary Search tree (BST) (by default over the end time attribute). This allows queries with predicates on either start or end time to be answered efficiently. It is heavily inspired by Priority Seach Trees (PST) ~\cite{CompGeometryBook}, a data structure traditionally used for storing intervals or 2D points in the context of computational geometry. In \DS, all queries are 3-sided (\autoref{fig:index-queries}), meaning that only one bound is specified for one of the dimensions. This allows temporal edges to be queried in $O(\log m + k)$, where $m$ is the number of temporal edges stored in it and $k$ is the number of results from the query. \autoref{fig:temporal_csr_and_TGER} shows a visual representation of \DS's data layout, contrasting it with T-CSR, and ~\autoref{algo:index-build} shows the pseudocode for \DS's parallel build operation.

\mypara{Ordering Predicates and 3-sided queries.} A key aspect of \DS is that the application can specify which of the two dimensions of the intervals should be mapped to the priority queue (or ``heap'') axis, and which dimension should be mapped to the ``BST'' axis. In addition, \DS can also be configured as either a min-heap, or a max-heap. Given this flexibility, both Succeeds and StrictlySucceeds ordering predicates can be translated to an equivalent 3-sided query by flipping the axis and/or using a max-heap. The Overlaps ordering predicate, however, requires both ends of the interval of two subsequent edges to be checked against each other, as can be seen in \autoref{fig:ordering_predicates}. In that case, \textbf{{\em TemporalEdgeMap}} (\autoref{ss:temporal_edgemap}) performs one additional query to match in-neighbors with corresponding out-neighbors.

\begin{algorithm}[!t]
\begin{algorithmic}[1]
\algnotext{EndFor}
\algnotext{EndParFor}
\Require{A temporal graph $G = (V, E)$.}
\Ensure{If applicable, the TGER for each vertex in $G$.}

\Procedure{IndexVertices}{$G$}
  \ParFor{each $v \in V$}
    \State $d = \text{out-degree of } v$  \Comment{Omitted: in-neighbors processing.}
    \If {$d \ge \text{min cutoff}$}  \Comment{(Section 5)}
        \State $\text{out-index}[v] =$ \Call{BuildIndex}{$e \in \text{out-edges of } v$}
    \EndIf
  \EndParFor
\EndProcedure

\Procedure{BuildIndex}{$E$}
  \State $\text{sorted} = \text{parallel-sort(E, ByStartTime())}$
  \State \Return \Call{BuildIndexRecurse}{$\text{sorted}, 0, \text{sorted.size()}$}
\EndProcedure

\Procedure{BuildIndexRecurse}{$\text{sorted, start, end}$}
  \State $\text{size = end - start}$
  \If {$\text{size} == 0$}
      \Return nullptr
  \EndIf
  \If {$\text{size} == 1$}
    \Return $\text{new TGERNode(sorted[start])}$
  \EndIf

  \State $\text{TGERNode* root = new TGERNode(sorted[start])}$
  \State start = start + 1

  \State mid-idx = index of point with median "end time" in sorted
  \State root->y-mid = sorted[mid-idx].end-time
  \State root->left = \textbf{spawn} \Call{BuildIndexRecurse}{$\text{sorted, start, mid-idx}$}
  \State root->right = \Call{BuildIndexRecurse}{$\text{sorted, mid-idx, end}$}
  \State \textbf{sync}  

\EndProcedure

\end{algorithmic}
\caption{Pseudocode for TGER index parallel build.}
\label{algo:index-build}
\end{algorithm}

\subsection{Temporal Edge Map}
\label{ss:temporal_edgemap}

We introduce \textbf{{\em TemporalEdgeMap}}, a programming primitive inspired by Ligra's \textbf{{\em EdgeMap}} (\autoref{ss:ligra_prim}), and designed to efficiently handle temporal information during parallel processing of temporal graphs. The \textsc{TemporalEdgeMap} extends Ligra's \textbf{{\em EdgeMap}} to the temporal setting by incorporating selective indexing (see\autoref{ss:cost_model}) to dynamically switch between different scanning strategies (Temporal CSR scan vs. \DS) depending on the query. Moreover, the primary distinction between \textbf{{\em TemporalEdgeMap}}'s programming interface and that of Ligra's \textbf{{\em EdgeMap}} is that it allows for {\em mapping over temporal edges while specifying an \textbf{ordering predicate}~(\autoref{ss:interface})} as input. Combined with \DS, the \textbf{{\em TemporalEdgeMap}} acts as a parallel mapping function that enables temporal graph algorithms to efficiently retrieve only the edges that satisfy temporal predicates of interest, while still respecting user-defined constraints regarding the validity of temporal paths in their applications. In~\autoref{algo:earliest-arrival}, we demonstrate the use of \textbf{{\em TemporalEdgeMap}} to update vertex frontiers for a parallel implementation of the earliest-arrival temporal path algorithm~\cite{TemporalPaths}. This example illustrates how the \textbf{{\em TemporalEdgeMap}} can be used to define complex suitable temporal graph processing logic in the context of parallel graph processing applications.

\begin{algorithm}[!t]
\begin{algorithmic}[1]
\Require{A temporal graph $G = (V, E)$, a target vertex $x \in V$, and a time interval $[t_a, t_b]$.}
\Ensure{$t[V]$: The earliest-arrival time from $x$ to every vertex $v \in V$ within query time interval $[t_a, t_b]$.}
\Procedure{Update}{$s, d, [t_s, t_e]$}
  \If {$t_s >= t_a$ \OR $t_e > t_b$}
    \Return 0
  \EndIf
  \If {$t_s < t[s]$ \OR $t_e >= t[d]$}
    \Return 0
  \EndIf
  \State \Return \Call{writeMin}{$t[d], t_e$} \AND \Call{CAS}{$\text{Visited}[d], 0, 1$}
\EndProcedure

\Procedure{Cond}{$i$}
  \State \Return $(\text{Visited}[i] == 0)$
\EndProcedure

\Procedure{EarliestArrival}{$G, x, [t_a, t_b]$}
  \State $t[x] = t_b$, $t[v] = \infty$ for all $v \in V \setminus \{x\}$
  \State $\text{Visited}[v] = 0$ for all $v \in V$ 

  \State $\text{Frontier} = \{x\}$
  \While{\Call{Size}{Frontier} $\neq 0$}
  \State Frontier $= $ \Call{TemporalEdgeMap}{$G, [t_a, t_b], \text{Frontier}$, \textproc{Update}, \textproc{Cond}, \textproc{OrdPred.StrictlySucceeds}}
  \EndWhile
\EndProcedure

\end{algorithmic}
\caption{Pseudocode for Earliest Arrival in Kairos.}
\label{algo:earliest-arrival}
\end{algorithm}

\section{Selective Indexing Optimization}
\label{s:indexing}

In this section, we introduce the concept of {\em selective indexing}. As we described \autoref{ss:overview-vertex_indexing}, the Vertex Indexer builds a \DS only for those vertices whose size (as out/in-degree) meets an experimentally obtained predefined vertex size threshold (as of writing, currently set to 2k edges). With selective indexing, we propose a novel cost model and related algorithms to {\em determine the most appropriate access method for each vertex in a temporal graph during traversal}. The unique aspect of the cost estimation problem we address is that it selectively assigns different access methods (e.g., index vs full scan) {\em for the same query} at runtime. As an example, while a conventional relational query optimizer may select an in-memory hash index for all primary keys satisfying a PrimaryKey-ForeignKey join, our selective indexing approach considers the estimated selectivity for the query based on the {\em value} of each primary key, as well as the anticipated number of matches it has in the corresponding foreign key table.

\begin{figure}[!t]
\centering\includegraphics[width=0.4\textwidth]{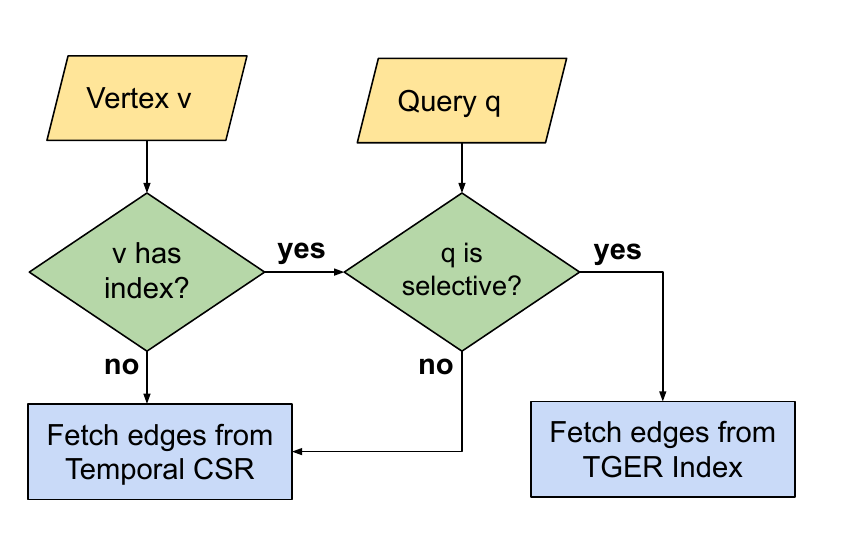}
\caption{Selective indexing: decision tree for accessing a vertex's neighboring edges given an input query.}
\label{fig:selective_indexing}
\end{figure}

\autoref{fig:selective_indexing} presents the decision tree used by \sysname's Vertex Indexer~(\autoref{fig:arch}) to determine the optimal candidate data structure for accessing a vertex's neighboring edges, given an input temporal query. The following sections provide a description of our selective indexing approach's cost model and associated algorithms.


\subsection{Cost Model}
\label{ss:cost_model}

To define a cost model for deciding when it is beneficial to use \DS compared to a parallel scan over the T-CSR, we need to consider the following factors:

\mypara{Vertex degree distribution.} The distribution of the number of outgoing edges for each vertex is important because \DS is only built (and potentially accessed) for vertices with more than a certain number of outgoing (or incoming) edges. If a large portion of the vertices have a high degree, the custom index will be more frequently accessed and contribute more to the overall query performance. Furthermore, more memory will also be used to store index data.

\mypara{Temporal edges distribution.} This refers to the distribution of start times and end times present in these edges for a given vertex. These distributions indicate the expected number of matches for a given temporal predicate query interval. Even though \DS is balanced, skew in the distribution of start times can lead to more levels being traversed to retrieve all edges that satisfy a given input predicate.

\mypara{Query workload.} For a given input temporal predicate, \sysname needs to estimate the expected number of results for a specific vertex corresponding to that query.\\

In summary, the distribution of the start and end times of the temporal edges affects the selectivity of the queries. If the distribution is such that the queries have a high degree of selectivity (i.e., they filter out a significant portion of the data), using the \DS will be more efficient. Conversely, if the distribution is such that the queries have low selectivity (i.e., they return a large portion of the data), \DS's performance advantage over the T-CSR access method is reduced.

Taking these factors into account, we define a cost model that estimates the time required for each method (index-based using \DS and T-CSR-based) to execute a given query. The model acts as a proxy for the estimated query processing time given what we know about the graph (e.g., cardinality of a vertex's neighborhood), the query workload (e.g., estimated selectivity of input query), and characteristics of the data structures used for access (e.g., parallelism potential and asymptotic complexity). For the \DS-based access method, the cost model relies on the time complexity for PSTs, which is $O(\log m + k)$, where $m$ is the number of elements stored in the data structure, and $k$ is the number of results matching the input query. A \DS is created for each vertex (\autoref{fig:selective_indexing}), so $m = deg(v)$, where $v \in V$, which yields

\begin{equation}
  T_v = c \cdot [\log(deg(v)) + k]
\end{equation}
where $c$ acts as a constant factor representing the average cost of performing a single operation using \DS. The value for $c$ captures the parallelism potential when using \DS, and is derived experimentally.

For the T-CSR access method, \sysname needs to perform a parallel scan to filter out edges that satisfy the input temporal predicate. Despite this operation being highly parallelizable and cache-friendly, the asymptotic time complexity for this scan is still $deg(v)$, as all edges for $v$ need to be scanned, with

\begin{equation}
S_v = c' \cdot deg(v)
\end{equation}
where $c'$ is a constant factor representing the average cost of performing a single operation using the T-CSR. It performs a role similar to that of $c$ for \DS, and like $c$, it too is derived experimentally. To decide which method is more beneficial for a given query, we compare the estimated time costs for both methods using the cost model. If the estimated time cost for \DS method ($T_v$) is lower than the estimated time cost for the T-CSR array method ($S_v$), then it is more beneficial to use \DS. Our cost model parameterizes the cardinality estimator based on the factors described above, with

\begin{equation}
C_v =
\begin{cases}
T_v & \text{if } \beta \leq \theta_{\text{sel}} \\
S_v & \text{otherwise} \\
\end{cases}
\end{equation}

\begin{conditions*}
C_v & the estimated cost of accessing a vertex's neighbors \\
T_v & the cost of querying for vertex $v$'s neighbors in \DS \\
S_v & the cost of scanning vertex $v$'s neighbors in T-CSR that satisfy the input temporal predicate\\
\beta & the selectivity of the input temporal predicate, defined as $k / m$ \\
\theta_{\text{sel}} & the selectivity threshold \\
\end{conditions*}

For example, assuming a selectivity threshold $\theta_{\text{sel}} = 0.3$ (i.e., queries retrieving 30\% of the neighboring edges of a vertex $v$), if the estimated selectivity $\beta$ is less than $0.3$ (e.g., $0.2$), then \sysname chooses \DS as the access method for vertex $v$ given input query $q$.

In theory, the choice of threshold primarily depends on the relationship between $k$ and $deg(v)$. As $k$ approaches $dev(v)$, the cost of using \DS converges to $O(deg(v))$, rather than $O(\log(deg(v)))$. In this case, it becomes more beneficial to use T-CSR due to its higher parallelism potential. Parallelism is higher for queries over T-CSR because they are implemented as highly parallelizable scans over a parallel array. On the other hand, \DS queries rely on divide-and-conquer recursive parallel operations over a data structure that resembles a BST in one dimension and a heap in another, leading to reduced parallelism behavior as $k$ nears $deg(v)$. In practice, however, we determine which threshold to use from experimental results, and find that 2k edges strikes a good performance vs estimator accuracy balance. By selecting an appropriate threshold, \sysname can decide at runtime whether to use the \DS index or T-CSR for accessing a vertex's neighbors, thereby improving query performance compared to a baseline~\cite{temporal_ligra} that only uses T-CSR.


\subsection{Cardinality Estimation Algorithm}
\label{ss:estimation_algo}


Our cardinality estimation algorithm aims to help determine the best access method for each vertex while taking into account the temporal predicates present in the query workload and the characteristics of the temporal graph being queried.

During the index construction phase (i.e., when building the \DS indices), \sysname creates a 2D density histogram for each vertex that meets the threshold for \DS index construction. The dimensions of the histogram are the start time and duration (end time $-$ start time) of the edges equally divided into $100$ buckets per dimension, for a total of $10000$ buckets, capturing the temporal distribution of a vertex's edges.
At runtime, \sysname uses the histogram to estimate the density of a vertex's edges that would satisfy the query's temporal predicates. Depending on the estimated density and a selectivity threshold, \sysname selects the most efficient access method. If the estimated density is above the threshold, the query execution for that vertex employs the associated \DS index; otherwise, a parallel scan on the T-CSR is performed. This enables \sysname to adapt the access method according to the unique features of the temporal graph and the specific temporal predicates in the query, leading to improved query performance, as demonstrated in~\autoref{s:experiments}.

\section{Experimental Results}
\label{s:experiments}

In this section, we present the results of an experimental evaluation of \sysname's query performance and scalability.

\mypara{Setup.} We use a 2nd Generation Intel\textsuperscript{\textregistered} Xeon\textsuperscript{\textregistered} Scalable Processor (Cascade Lake) system with 24 physical (48 virtual) cores on each of its two NUMA nodes.  The socket has a total of 192GiB of DDR4 2666 MhZ RAM. Our programs use Cilk Plus~\cite{cilkplus} and are compiled with OpenCilk's~\cite{opencilk} clang version 14 and -O3 flag. All of the parallel speedup numbers that we report are based on the running time on 24-cores (single socket) without hyper-threading compared to the running time on a single thread.



\mypara{Datasets.} \autoref{tab:datasets} shows the number of vertices ($|V|$), number of edges ($|E|$), maximum out-degree ($\max_{v \in V} \textit{deg}^{+}(v)$), maximum in-degree ($\max_{v \in V} \textit{deg}^{-}(v)$), and average degree ($\textit{avg}_{v \in V} \textit{deg}(v)$) for each of the datasets under use. The \textsc{synthetic} data set comprises a temporal graph in which vertices are log-normally distributed, the inter-arrival times of start times follow a Poisson distribution, and the edge durations follow an uniform distribution. If the temporal edges in a dataset only have start times, then end time is sampled from a uniform distribution, similar to what is done in~\cite{TemporalPaths, TemporalPathsJournal}.

\begin{table}[htb!]
  \begin{center}

\resizebox{0.5\textwidth}{!}{%

    \begin{tabular}{l|c|c|c|c|cc}
      \toprule
        Name & $|V|$ & $|E|$ 
        & $\max_{v \in V} \textit{deg}^{+}(v)$ 
        & $\max_{v \in V} \textit{deg}^{-}(v)$ 
        & $\textit{avg}_{v \in V} \textit{deg}(v)$ \\
      \midrule
      bitcoin \cite{TemporalMotifsSampling}
        & $4.80 \times 10^7$
        & $1.13 \times 10^8$
        & $2.66 \times 10^6$
        & $2.53 \times 10^6$
        & $4$ \\
      netflow \cite{Dataset_netflow}
        & $3.72 \times 10^8$
        & $1.20 \times 10^9$
        & $5.78 \times 10^5$
        & $1.82 \times 10^8$
        & $12$ \\
      reddit-reply \cite{TemporalMotifsSampling}
        & $1.17 \times 10^7$
        & $6.46 \times 10^8$
        & $3.92 \times 10^5$
        & $9.94 \times 10^5$
        & $110$ \\      
      stackoverflow \cite{Dataset_stackoverflow}
        & $6.02 \times 10^6$
        & $6.34 \times 10^7$
        & $1.01 \times 10^5$
        & $93143$
        & $20$ \\
      transportation \cite{Dataset_transportation}
        & $41794$
        & $7.93 \times 10^7$
        & $50881$
        & $50625$
        & $3798$ \\      
      twitter-cache \cite{Dataset_twittercache}
        & $94226$
        & $8.55 \times 10^8$
        & $5.21 \times 10^6$
        & $2.31 \times 10^5$
        & $36328$ \\
      synthetic 
        & $10^7$
        & $10^9$
        & $5.65 \times 10^7$
        & $3.08 \times 10^7$
        & $99$ \\
      \bottomrule
    \end{tabular}

  } 

  \end{center}
  \caption{Temporal graphs used for evaluation}
  \label{tab:datasets}  
\end{table}


\begin{figure}[htb!]
\centering\includegraphics[width=0.4\textwidth]{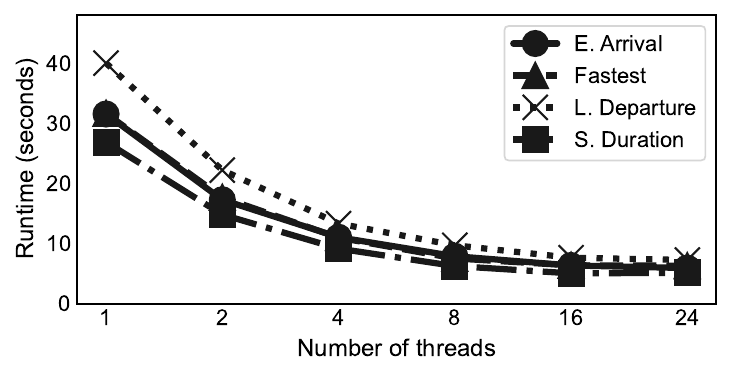}
\caption{Running time vs number of threads on a smaller synthetic dataset (500M edges).}
\label{fig:speedup}
\end{figure}


\subsection{Scalability}
\label{ss:exps_scale}

\autoref{tab:app_speedup} shows the sequential and parallel running times of our algorithms, as well as their parallel speedup. Runtimes are reported as an average of ten runs. Except for T. BC, T. CC, k-core, and PageRank, all algorithm runtimes use a single source vertex. For algorithms requiring a single source, we selected the top 100 vertices based on their out-degree, leading to 100 runs in a single execution. For PageRank, the reported runtimes cover 100 iterations. T. BC uses the number of temporal S. Duration paths when calculating centrality for each vertex. The algorithms E. Arrival, L. Departure, Fastest, and S. Duration inherently expect a start and end time in their original definitions. For BC, BFS, CC, $k$-core, and PageRank, we have adapted the original algorithms to accept a start and end time as input. For every algorithm, we set the start time to align with the 95th percentile of the latest start times in the dataset. The end time is set to the maximum value, representing the 100th percentile.

The algorithms overall get good parallel speedup, with a maximum of 22.6x and mean speedup 8.7x. The lower speedups are in general observed in the smaller datasets, as they are too small to benefit from parallel processing. However, the larger speedups are not necessarily always observed in the largest dataset. Rather, they seem to be influenced by a number of factors, including how skewed the vertex degree distribution is, as well as fraction of edges matching the input temporal predicate. In other words, while we use the same query interval size for all experiments, the edges matching that input query predicate (i.e., its selectivity) may not be evenly distributed over all vertices, and thus do not offer the same parallelism potential. Furthermore, some co-routines in \sysname are only parallelized if the minimum number of interations meets an heuristic threshold (e.g., at least 1000 iterations in parallel \textsc{cilk\_for} loops), which again may be influenced by skew in the underlying vertex-degree and inter-arrival times distribution for a given dataset.

\autoref{fig:speedup} shows the running time vs.\ number of threads for all of the minimal temporal paths over a smaller synthetically generated dataset (500M edges). We see good parallel scalability for all of these algorithms, with speedups ranging from 5x to 8x, though see little benefit from 16 to 24 cores.


\begin{figure}[htb!]
\centering\includegraphics[width=0.4\textwidth]{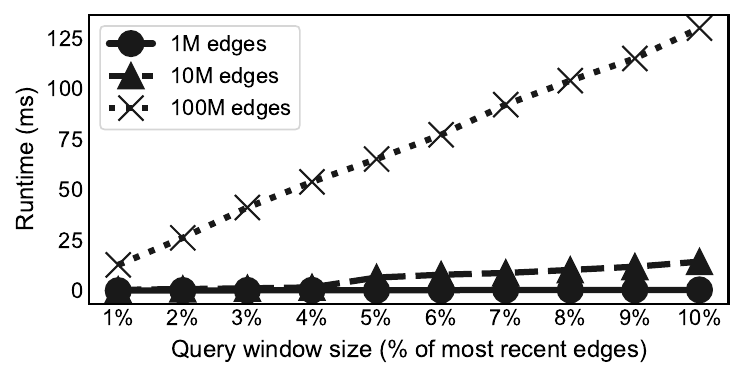}
\caption{Runtimes vs.\ query window size (fraction of most recent edges) for different TGER sizes.}
\label{fig:microbench_query}
\end{figure}


\subsection{Impact of Selective Indexing}
\label{ss:exps_perf}

\begin{figure*}[t]
  \begin{center}
  \includegraphics[width=0.85\linewidth]{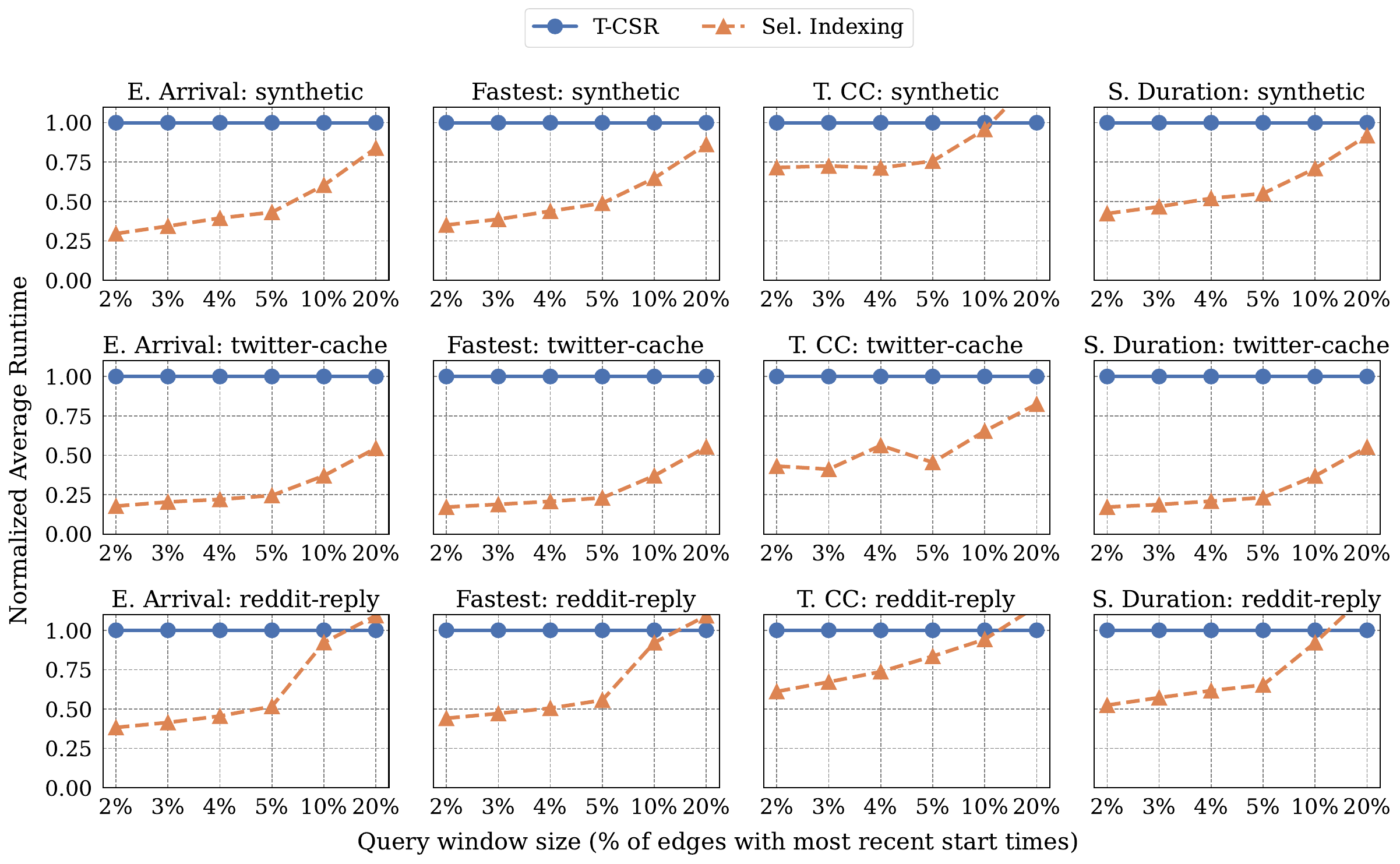}
  \caption{Normalized average runtime vs input query window size. Boundaries of the input query are varied so as to match exactly against a fixed percentage of most recent edges present in the corresponding temporal graph dataset. Runtimes are normalized against T-CSR as baseline~\cite{temporal_ligra}.}
  \label{fig:tcsr_vs_selindex}
  \end{center}
\end{figure*}

\autoref{fig:tcsr_vs_selindex} shows a comparison against a Temporal Ligra baseline~\cite{temporal_ligra}, i.e., T-CSR is used for all vertices and no selective indexing is present, as described in \autoref{ss:temporal_csr}. As far as we know, this is the fastest available shared-memory implementation of graph processing algorithms (temporal and traditional) over temporal graphs. The algorithms are configured in the same manner as in \autoref{ss:exps_scale}, and here we show the running times for a subset comprised mostly of temporal minimal paths, and over one small (reddit-reply), and two large datasets (twitter-cache and synthetic). Our results show that the selective indexing approach does reasonably well, with up to 8x improvement in some cases. As expected, the highly selective queries are the ones with the most improvement. Furthermore, between 10\% and 20\% selectivity, the T-CSR approach starts being more advantageous. 


\subsection{Microbenchmark: \DS query runtimes}
\label{ss:microbench}

\autoref{fig:microbench_query} shows the time it takes to run a query over a single \DS of various sizes (synthetically generated 1M, 10M and 100M edges). We note that each \DS in \sysname is associated to a single vertex. In other words, the running times reported in this experiment should be interpreted as the amount it takes to retrieve edges of interest from a {\em  single vertex} that has been indexed with \DS. Similar to the results reported in experiments above, input queries are sized to match a portion of the most recent edges (by start time) in the input data (in this case, a vertex's neighboring edges). Results here show that it takes less than 125 milliseconds to retrieve roughly 10\% of a \DS with 100M edges.

\subsection{Comparison to Alternatives}
\label{ss:exps_qual}

\mypara{Distributed memory.} Although a direct comparison with Tink, ICM is challenging due to their distributed memory design, we offer a comparison based on the runtimes presented in their papers~\cite{Tink, ICM}. Using 8 cores, Tink processes the E. Arrival algorithm for just one source vertex in 37s on a synthetic graph of 25 million edges. In contrast, \sysname, with the same number of cores, processes the same algorithm for the 100 source vertices (equivalent to 100 runs) on a synthetic dataset of 500M edges, as shown in~\autoref{fig:speedup} in less than 10s. {\em Put simply, for E. Arrival -- the sole algorithm with reported runtimes by Tink -- \sysname handles data that's 20 times larger, addresses 100 times more source vertices, and completes the task in a third of the time}. We could not locate total runtimes for ICM in its publication. However, in its ``weak scaling'' analysis, it reports nearly 500 seconds to run E. Arrival on a single source vertex for a graph segment with 100M edges, using 8 cores. \sysname, in contrast, requires under 10 seconds for 100 vertices on 500 million edge dataset (\autoref{fig:speedup}). {\em In essence, when compared with ICM, \sysname operates 50 times faster, handles 100 times more algorithm executions, processing a dataset that is 50 times larger}.

\mypara{Shared memory.} At the time of writing, the only other temporal graph analytics system in this category is TeGraph~\cite{TeGraph}. We have contacted the authors, who shared an implementation of shortest paths with us, as well \textsc{Delicious}, one of the datasets they used for evaluation. This dataset has around 300M edges, and 34M vertices. Unfortunately, we were not able to reproduce the results in~\cite{TeGraph}, which report around one second as the total runtime for executing shortest path on the aforementioned dataset using 100 source vertices. Instead, we find that their implementation of shortest path takes around three seconds to process a single vertex, regardless of which vertex is provided as input. Extrapolating from this result, we get around 300 seconds to process 100 source vertices -- as opposed to the one second reported in their paper. Furthermore, the results for their ``OnePass'' baseline (which is closer to our approach) are significantly slower than \sysname. Specifically, for the same \textsc{Delicious} dataset, their ``OnePass'' baseline takes close to 90 seconds for 100 vertices using 16 threads on a 8-core machine. {\em \sysname, on the other hand, takes around five seconds using 8 cores, which is a 18x speedup compared to their ``OnePass'' baseline, and 60x compared to our local results from their implementation.}

\begin{table*}[t!]
\begin{center}
\setlength\tabcolsep{1.5pt} 

\resizebox{0.8\textwidth}{!}{%

\begin{tabular}{
@{} l @{\hspace{0.5\tabcolsep}}
c @{\hspace{0.5\tabcolsep}} |c| @{\hspace{0.5\tabcolsep}} c @{\hspace{0.5\tabcolsep}} ||
c @{\hspace{0.5\tabcolsep}} |c| @{\hspace{0.5\tabcolsep}} c @{\hspace{0.5\tabcolsep}} ||
c @{\hspace{0.5\tabcolsep}} |c| @{\hspace{0.5\tabcolsep}} c @{\hspace{0.5\tabcolsep}} ||
c @{\hspace{0.5\tabcolsep}} |c| @{\hspace{0.5\tabcolsep}} c @{\hspace{0.5\tabcolsep}} ||
c @{\hspace{0.5\tabcolsep}} |c| @{\hspace{0.5\tabcolsep}} c @{\hspace{0.5\tabcolsep}} ||
c @{\hspace{0.5\tabcolsep}} |c| @{\hspace{0.5\tabcolsep}} c @{\hspace{0.5\tabcolsep}} ||
c @{\hspace{0.5\tabcolsep}} |c| @{\hspace{0.5\tabcolsep}} l @{} }\toprule

& \multicolumn{3}{c}{bitcoin}
& \multicolumn{3}{c}{netflow}
& \multicolumn{3}{c}{reddit-reply}
& \multicolumn{3}{c}{stackoverflow}
& \multicolumn{3}{c}{transportation}
& \multicolumn{3}{c}{twitter-cache}
& \multicolumn{3}{c}{synthetic}
\\
\cmidrule(lr){2-4}
\cmidrule(lr){5-7}
\cmidrule(lr){8-10}
\cmidrule(lr){11-13}
\cmidrule(lr){14-16}
\cmidrule(lr){17-19}
\cmidrule(lr){20-22}

Application
& $T_{1}$ & $T_{24}$ & SU  
& $T_{1}$ & $T_{24}$ & SU  
& $T_{1}$ & $T_{24}$ & SU  
& $T_{1}$ & $T_{24}$ & SU  
& $T_{1}$ & $T_{24}$ & SU  
& $T_{1}$ & $T_{24}$ & SU  
& $T_{1}$ & $T_{24}$ & SU  
\\

\midrule
E. Arrival 
 & 74.2 
 & 7.87 
 & 9.4 
 & 124.8 
 & 21.6 
 & 5.8 
 & 3.89 
 & .76 
 & 5.1 
 & 17.1 
 & 1.99 
 & 8.6 
 & 58.8 
 & 6.49 
 & 9.1 
 & 246.9 
 & 16.11 
 & 15.3 
 & 440 
 & 31.7
 & 13.9 \\ 
L. Departure
 & 56.4 
 & 6.63 
 & 8.5 
 & 125.2 
 & 22 
 & 5.7 
 & 5.1 
 & .76 
 & 3.9 
 & 11.2
 & 1.53 
 & 7.3 
 & 45.4
 & 5.22
 & 8.7
 & 214.2 
 & 15.15
 & 14.1
 & 391
 & 30.6
 & 12.8\\ 
Fastest         
 & 56.3
 & 7.73
 & 7.3
 & 185.6
 & 31.6
 & 5.9
 & 5.81
 & 1.11
 & 5.2
 & 9.16
 & 1.32
 & 6.9
 & 38.7
 & 4.73
 & 8.2
 & 222.6
 & 12.84
 & 17.4
 & 311
 & 23.1
 & 13.5\\ 
S. Duration
 & 56.6
 & 7.76
 & 7.3
 & 187
 & 32.2
 & 5.8
 & 5.84
 & 1.07
 & 5.5
 & 9.22
 & 1.33
 & 6.9
 & 2.7
 & 19.7
 & 7.3
 & 12.84
 & 223.5
 & 17.4
 & 285
 & 21.6
 & 13.2\\ 
\midrule
T. BFS       
 & 8.5
 & 1.5
 & 5.7
 & 97.2
 & 16.5
 & 5.9
 & 1.9
 & .4
 & 4.8
 & .8
 & .2
 & 4
 & .009
 & .002
 & 4.5
 & 9.8
 & 1.4
 & 7
 & 19.6
 & 3.2
 & 6.2\\ 
T. CC        
 & 5.65
 & .59
 & 9.6
 & 24.4
 & 2.2
 & 11.1
 & 3.99
 & .2
 & 19.9
 & 1.85
 & .11
 & 16.8
 & 6.83
 & .49
 & 13.9
 & 10.2
 & .71
 & 14.4
 & 5.41
 & 4.57
 & 1.2\\ 
T. k-core    
 & 13.7
 & 1.22
 & 11.4
 & 38.9
 & 3.6
 & 10.8
 & 32.4
 & 4.2
 & 7.7
 & 2.2
 & .4
 & 5.5
 & .02
 & .01
 & 2
 & 2.7
 & .9
 & 3
 & 2
 & .5
 & 4\\ 
\midrule
T. BC        
 & 10.8
 & .57
 & 18.95
 & 8.04
 & 1.26
 & 6.38
 & .032
 & .168
 & 5.25
 & .085
 & .014
 & 6.07
 & .005
 & .003
 & 1.67
 & 2.84
 & .692
 & 4.1
 & 4.57
 & 3.76
 & 1.22\\ 
T. PageRank  
 & 65.3
 & 5.23
 & 12.5
 & 35.8
 & 2.6
 & 13.8
 & 11.3
 & .50
 & 22.6
 & 8.51
 & .42
 & 20.3
 & 2.6
 & .21
 & 12.4
 & 17.22
 & 1.2
 & 14.4
 & 11
 & 4.62
 & 2.4\\ 
\bottomrule
    \end{tabular}
} 

  \end{center}
  \caption{Running times (in seconds) of single-threaded ($T_{1}$), 24-core no hyper-threading ($T_{24}$), and parallel speedup as single-thread time divided by 24-core time (SU).}
  \label{tab:app_speedup}
\end{table*}

\subsection{Selective Indexing: Estimator Accuracy}
\label{ss:exps_indexing_accuracy}

In \autoref{ss:cost_model}, we proposed a novel approach for selectively indexing a subset of vertices in a temporal graph, as well as cost model for deciding which access method to use for each vertex at runtime. In this section, we present an experiment assessing the accuracy of the {\em cardinality estimator}, a key component of our proposed cost model. For this evaluation, we vary the size of an input temporal predicate, and use a selectivity threshold of 20\% (the ``q is selective?'' decision in \autoref{fig:selective_indexing}). As in the experiments we describe above, here we also vary the size of an input query interval to purposefully match against a percentage of the most recent edges (by start time) in the dataset. To measure accuracy, we define true positives as ``should use \DS, and did'' and true negatives as ``should not use \DS, and did not''. Specifically, ``should'' here takes into account the estimated selectivity compared to an oracle with the actual selectivity of the query. Furthermore, we only evaluate this decision for vertices that have been large enough to be indexed with a \DS (otherwise, no decision needs to be made, as T-CSR is used a 100\% of the time), and vary the minimum cutoff for indexing from 1k to 8k edges (multiples of 2). We find that for all datasets, the accuracy of this decision stays consistently above 90\% for input query intervals sized under 1\%, and above 95\% for all other input query interval sizes we assessed (2 through 5\%, 10\%, and 20\%). Furthermore, the accuracy also increases as a function the cutoff size. As expected, this increase is largely due to the cardinality estimator's 2D density histogram (\autoref{ss:estimation_algo}) having more samples in this case.


\section{Related Work}
\label{s:related}

\mypara{Temporal Graph Analytics.} While the authors of~\cite{TemporalPaths, TemporalPathsJournal} have been the first to propose one-pass parallel versions of temporal graph algorithms, as far as we know~\cite{TeGraph} is the only other shared-memory temporal graph analytics system. Previous systems for temporal graph analytics have primarily relied on distributed message passing, often building upon programming paradigms like Pregel or stateful data stream processing frameworks such as Apache Flink~\cite{ICM, ICMEngine, Tink, GRADOOP}. The large number of messages exchanged when relying on these programming paradigms imposes considerable overhead on graph processing. This becomes particularly noticeable in the case of graphs that fit within available memory of current commodity servers, as shown by the speedups we get when compared against these systems.

\mypara{Time-evolving Graph Engines.} There has been significant work on developing frameworks for processing of graphs that evolve over time~\cite{Chronos, ImmortalGraph, Kineograph, graphchi, GraphTau}. While these systems allow processing of snapshots, streaming graphs, or dynamic graphs dynamic graphs, for their most part they do not support temporal graphs.  Rather, timestamps associated to edges or vertices are treated as graph updates in these systems. For this reason, we consider these orthogonal.

\mypara{Range Query Data Structures.} There has been extensive work on data structures to handle range queries. Traditional range query data structures often used in relational systems include quad-trees, R-trees, kd-trees. Specifically aimed at interval data -- such as that present in temporal edges -- are interval trees, segment trees, and priority search trees~\cite{CompGeometryBook}. All three of these data structures have runtime complexity of $O(log n + k)$, where $k$ is the number of results for the range query. Where they differ is on space complexity, which is highest for segment trees at $O(n log n)$, as well as on properties relating to how they can be queried (e.g., stabbing vs 3-sided queries). The index for storing temporal edges that we introduce in this paper is a highly-optimized and specialized version of a priority search tree.

\mypara{GNNs.} Graph Neural Networks (GNNs) have emerged as a powerful framework for learning representations of graph-structured data, tackling problems in various domains such as social networks, molecular biology, and recommendation systems~\cite{DGL,TGL,GNNs_temporal}. GNNs rely on the underlying graph structure and the attributes of nodes and edges to learn complex patterns and make predictions, with recent increased focus on temporal graphs~\cite{TGL, TGLBenchmark}. By employing efficient graph analytics algorithms, these systems can compute a wide range of graph properties, such as centrality measures, graph motifs, or community structures, which can then be used as additional input features for GNNs. In the context of GNNs, most related to our work is the version of Temporal CSR used for sampling temporal edges in~\cite{TGL}. While it does not account for end time edges, as far as we know this is the only other work that extends CSR to the temporal setting.

\section{Conclusions}
\label{s:conclusions}

We presented \sysname, a temporal graph analytics system that provides application developers a framework for efficiently executing temporal algorithms over temporal graphs. Specifically, it employs \DS, a highly-optimized parallel data structure, as efficient index for temporal graph processing. \sysname is built atop Ligra~\cite{ligra}, a state-of-the-art parallel graph processing system. With \DS, Ligra's vertex-centric computational model takes advantage of locality that is naturally occurring on temporal graphs and queries over such graphs. We show in our experiments that using \sysname, a number of minimal temporal path algorithms are up to 8x times faster than an already competitive baseline~\cite{temporal_ligra}, which we also provide. When compared with alternative shared-memory system, it achieves up to 60x speedups.



\begin{acks}
This research is supported by DOE Early Career Award \#DE-SC0018947, NSF CAREER
Award \#CCF-1845763, and by Intel as part of the MIT Data Systems and AI
Lab (DSAIL) at MIT. Joana M. F. da Trindade was partially supported by an
Alfred P. Sloan UCEM PhD Fellowship, and a Microsoft Research PhD Fellowship.
\end{acks}


\emergencystretch=1em
\bibstyle{ACM-Reference-Format}
\printbibliography

@ARTICLE{TGLBenchmark,
  title         = "Temporal Graph Benchmark for Machine Learning on Temporal
                   Graphs",
  author        = "Huang, Shenyang and Poursafaei, Farimah and Danovitch, Jacob
                   and Fey, Matthias and Hu, Weihua and Rossi, Emanuele and
                   Leskovec, Jure and Bronstein, Michael and Rabusseau,
                   Guillaume and Rabbany, Reihaneh",
  year          =  2023,
  archivePrefix = "arXiv",
  primaryClass  = "cs.LG",
  eprint        = "2307.01026"
}

@inproceedings{Dataset_twittercache,
  author = {Juncheng Yang and Yao Yue and K. V. Rashmi},
  title = {A large scale analysis of hundreds of in-memory cache clusters at Twitter},
  booktitle = {OSDI},
  year = {2020},
  pages = {191--208},
}

@ARTICLE{Dataset_transportation,
  title    = "A collection of public transport network data sets for 25 cities",
  author   = "Kujala, Rainer and Weckstr{\"o}m, Christoffer and Darst, Richard
              K and Mladenovi{\'c}, Milo{\v s} N and Saram{\"a}ki, Jari",
  journal  = "Nature Sci. Data",
  volume   =  5,
  year     =  2018,
}

@inproceedings{TemporalMotifsSampling,
  title={Sampling methods for counting temporal motifs},
  author={Paul Liu and Austin R. Benson and Moses Charikar},
  booktitle={WSDM},
  year={2019}
}

@inbook{Dataset_netflow,
  author = {Melissa J. M. Turcotte and Alexander D. Kent and Curtis Hash},
  title = {Unified Host and Network Data Set},
  booktitle = {Data Science for Cyber-Security},
  pages = {1-22},
  year = {2018},
}

@INPROCEEDINGS{Motifs1,
  title     = "Motifs in Temporal Networks",
  author    = "Paranjape, Ashwin and Benson, Austin R and Leskovec, Jure",
  pages     = "601--610",
  booktitle    = "WSDM",
  year      =  2017,
}

@ARTICLE{Motifs2,
  title    = "From networks to optimal higher-order models of complex systems",
  author   = "Lambiotte, Renaud and Rosvall, Martin and Scholtes, Ingo",
  journal  = "Nature Phys.",
  volume   =  15,
  number   =  4,
  pages    = "313--320",
  year     =  2019,
}

@ARTICLE{Epi1,
  title    = "Use of temporal contact graphs to understand the evolution of
              {COVID-19} through contact tracing data",
  author   = "Wu, Mincheng and Li, Chao and Shen, Zhangchong and He, Shibo and
              Tang, Lingling and Zheng, Jie and Fang, Yi and Li, Kehan and
              Cheng, Yanggang and Shi, Zhiguo and Sheng, Guoping and Liu, Yu
              and Zhu, Jinxing and Ye, Xinjiang and Chen, Jinlai and Chen,
              Wenrong and Li, Lanjuan and Sun, Youxian and Chen, Jiming",
  journal  = "Nature Commun. Phys",
  volume   =  5,
  year     =  2022,
}

@ARTICLE{Indoor1,
  title     = "{ASTRO}: Reducing {COVID-19} Exposure through Contact Prediction
               and Avoidance",
  author    = "Anastasiou, Chrysovalantis and Costa, Constantinos and
               Chrysanthis, Panos K and Shahabi, Cyrus and Zeinalipour-Yazti,
               Demetrios",
  journal   = "ACM Trans. Spatial Algorithms Syst.",
  volume    =  8,
  number    =  2,
  pages     = "1--31",
  year      =  2022,
}

@article{Indoor2,
  author = {Liu, Tiantian and Li, Huan and Lu, Hua and Cheema, Muhammad Aamir and Shou, Lidan},
  title = {Towards Crowd-Aware Indoor Path Planning},
  year = {2021},
  pages = {1365–1377},
  journal = {PVLDB}
}

@inproceedings{Provenance2,
  title={You Are What You Do: Hunting Stealthy Malware via Data Provenance Analysis.},
  author={Wang, Qi and Hassan, Wajih Ul and Li, Ding and Jee, Kangkook and Yu, Xiao and Zou, Kexuan and Rhee, Junghwan and Chen, Zhengzhang and Cheng, Wei and Gunter, Carl A and others},
  booktitle={NDSS},
  year={2020}
}

@inproceedings{Provenance1,
author = {Erb, Benjamin and Mei\ss{}ner, Dominik and Pietron, Jakob and Kargl, Frank},
title = {Chronograph: A Distributed Processing Platform for Online and Batch Computations on Event-Sourced Graphs},
booktitle = {DEBS},
pages = {78--87},
year = {2017}
}

@inproceedings{Minimal1,
 author={Li, Lei and Wang, Sibo and Zhou, Xiaofang},
 title={Time-Dependent Hop Labeling on Road Network},
 booktitle={ICDE},
 year={2019},
 pages={902--913},
}

@article{Centrality1,
  title={Relevance of temporal cores for epidemic spread in temporal networks},
  author={Ciaperoni, Martino and Galimberti, Edoardo and Bonchi, Francesco and Cattuto, Ciro and Gullo, Francesco and Barrat, Alain},
  journal={Scientific reports},
  volume={10},
  number={1},
  pages={12529},
  year={2020},
  publisher={Nature Publishing Group UK London}
}

@inproceedings{Clusters4,
 title={Persistent community search in temporal networks},
 author={Li, Rong-Hua and Su, Jiao and Qin, Lu and Yu, Jeffrey Xu and Dai, Qiangqiang},
 booktitle={ICDE},
 year={2018},
 pages={797--808},
}

@article{Clusters3,
  title={On querying historical k-cores},
  author={Yu, Michael and Wen, Dong and Qin, Lu and Zhang, Ying and Zhang, Wenjie and Lin, Xuemin},
  year={2021},
% booktitle = {VLDB 2021},
 journal = {PVLDB},
 pages = {2033--2045},
}

@article{Clusters2,
  title={Scalable Time-Range k-Core Query on Temporal Graphs (Full Version)},
  author={Yang, Junyong and Zhong, Ming and Zhu, Yuanyuan and Qian, Tieyun and Liu, Mengchi and Yu, Jeffery Xu},
  year={2023},
% booktitle = {VLDB 2023},
 journal = {PVLDB},
 pages = {1168--1180},
}

@article{Clusters1,
  title={Mining Bursting Core in Large Temporal Graphs},
  author={Qin, Hongchao and Li, Rong-Hua and Yuan, Ye and Wang, Guoren and Qin, Lu and Zhang, Zhiwei},
  year={2022},
% booktitle = {VLDB 2022},
 journal = {PVLDB},
 pages = {3911--3923},
}

@article{Reachability1,
  title={Efficient distributed reachability querying of massive temporal graphs},
  author={Zhang, Tianming and Gao, Yunjun and Chen, Lu and Guo, Wei and Pu, Shiliang and Zheng, Baihua and Jensen, Christian S},
% booktitle = {VLDB 2019}
  journal={PVLDB},
  pages={871--896},
  year={2019},
}

@article{InfoPropagation2,
  title={Universality, criticality and complexity of information propagation in social media},
  author={Notarmuzi, Daniele and Castellano, Claudio and Flammini, Alessandro and Mazzilli, Dario and Radicchi, Filippo},
  journal={Nature communications},
  volume={13},
  number={1},
  pages={1308},
  year={2022},
  publisher={Nature Publishing Group UK London}
}

@article{InfoPropagation1,
  title={Uncovering the structure and temporal dynamics of information propagation},
  author={Manuel Gomez Rodriguez and Jure Leskovec and David Balduzzi and Bernhard Schölkopf},
  journal={Network Science},
  volume={2},
  number={1},
  pages={26 - 65},
  year={2014},
  publisher={Cambridge University Press},
}

@INPROCEEDINGS{leskovec2005graphs,
  title     = "Graphs over time: densification laws, shrinking diameters and
               possible explanations",
  booktitle = "KDD",
  author    = "Leskovec, Jure and Kleinberg, Jon and Faloutsos, Christos",
  pages     = "177--187",
  series    = "KDD",
  year      =  2005,
}

@ARTICLE{DGL,
  title    = "Deep Graph Library: towards efficient and scalable deep learning
              on graphs",
  author   = "Wang, Minjie Yu",
  journal  = "ICLR",
  year     =  2019
}

@article{TGL,
  author = {Zhou, Hongkuan and Zheng, Da and Nisa, Israt and Ioannidis, Vasileios and Song, Xiang and Karypis, George},
  title = {TGL: A General Framework for Temporal GNN Training on Billion-Scale Graphs},
  year = {2022},
  journal = {PVLDB},
  pages = {1572–1580},
}

@article{GNNs_temporal,
  title={Graph Neural Networks for Temporal Graphs: State of the Art, Open Challenges, and Opportunities},
  author={Antonio Longa and Veronica Lachi and Gabriele Santin and Monica Bianchini and Bruno Lepri and Pietro Lio and franco scarselli and Andrea Passerini},
  journal={Transactions on Machine Learning Research},
  year={2023},
  url={https://openreview.net/forum?id=pHCdMat0gI},
}

@ARTICLE{interval_algebra,
  title     = "Maintaining knowledge about temporal intervals",
  author    = "Allen, James F",
  journal   = "Commun. ACM",
  volume    =  26,
  number    =  11,
  pages     = "832--843",
  year      =  1983,
}

@article{ICMEngine,
  title={A Distributed Path Query Engine for Temporal Property Graphs},
  author={Shriram Ramesh and Animesh Baranawal and Yogesh Simmhan},
  journal   = {CoRR 2020},
  volume    = {abs/2002.03274},
  NOTE      = {{P}reprint, \url{https://arxiv.org/abs/2002.03274}},
  archivePrefix = {arXiv}
}

@inproceedings{ICM,
 author={{S. Gandhi and Y. Simmhan}},
 title={An interval-centric model for distributed computing over temporal graphs},
 booktitle={ICDE},
 year={2020},
 pages={1129--1140}
}

@inproceedings{Tink,
author = {Lightenberg, Wouter and Pei, Yulong and Fletcher, George and Pechenizkiy, Mykola},
title = {Tink: A Temporal Graph Analytics Library for Apache Flink},
year = {2018},
booktitle = {WWW},
pages = {71–72},
}

@article{GRADOOP,
  title    = "Distributed temporal graph analytics with {GRADOOP}",
  author   = "Rost, Christopher and Gomez, Kevin and T{\"a}schner, Matthias and
              Fritzsche, Philip and Schons, Lucas and Christ, Lukas and
              Adameit, Timo and Junghanns, Martin and Rahm, Erhard",
  %booktitle = {PVLDB 2022},
  journal = {PVDLB},
  pages    = {375--401},
  year     =  {2022},
}

@INPROCEEDINGS{TeGraph,
  title     = "{TeGraph}: A Novel {General-Purpose} Temporal Graph Computing
               Engine",
  booktitle = {ICDE},
  author    = "Huan, Chengying and Liu, Hang and Liu, Mengxing and Liu,
               Yongchao and He, Changhua and Chen, Kang and Jiang, Jinlei and
               Wu, Yongwei and Song, Shuaiwen Leon",
  pages     = "578--592",
  year      =  2022,
}

@inproceedings{Kineograph,
  author    = {Raymond Cheng and
               Ji Hong and
               Aapo Kyrola and
               Youshan Miao and
               Xuetian Weng and
               Ming Wu and
               Fan Yang and
               Lidong Zhou and
               Feng Zhao and
               Enhong Chen},
  title     = {Kineograph: taking the pulse of a fast-changing and connected world},
%  booktitle = {EuroSys 2012},
  booktitle = {EuroSys},
  year = {2012},
  pages     = {85--98},
}

@article{ImmortalGraph,
  author    = {Youshan Miao and
               Wentao Han and
               Kaiwei Li and
               Ming Wu and
               Fan Yang and
               Lidong Zhou and
               Vijayan Prabhakaran and
               Enhong Chen and
               Wenguang Chen},
  title     = {ImmortalGraph: {A} System for Storage and Analysis of Temporal Graphs},
  journal   = {{TOS}},
  volume    = {11},
  number    = {3},
  pages     = {14:1--14:34},
  year      = {2015},
}

@inproceedings{GraphTau,
 author = {Iyer, Anand Padmanabha and Li, Li Erran and Das, Tathagata and Stoica, Ion},
 title = {Time-evolving Graph Processing at Scale},
 %booktitle = {GRADES 2016},
 booktitle = {GRADES},
 year = {2016},
 pages = {5:1--5:6},
}

@inproceedings{Chronos,
 author = {Han, Wentao and Miao, Youshan and Li, Kaiwei and Wu, Ming and Yang, Fan and Zhou, Lidong and Prabhakaran, Vijayan and Chen, Wenguang and Chen, Enhong},
 title = {Chronos: A Graph Engine for Temporal Graph Analysis},
 booktitle = {EuroSys},
 year = {2014},
 pages = {1:1--1:14},
}

@book{CompGeometryBook,
 author = {Berg, Mark de and Cheong, Otfried and Kreveld, Marc van and Overmars, Mark},
 title = {Computational Geometry: Algorithms and Applications},
 year = {2008},
 edition = {3rd ed.},
 publisher = {Springer-Verlag TELOS},
}

@article{TemporalNetworks,
 author = {Holme, Petter and Saram\"{a}ki, Jari},
 journal = {Physics Reports},
 pages = {97--125},
 title = {{Temporal networks}},
 url = {http://arxiv.org/abs/1108.1780},
 volume = {519},
 year = {2012}
}

@ARTICLE{AdaptingToSkew,
  title         = "Adapting to Skew: Imputing Spatiotemporal Urban Data with
                   {3D} Partial Convolutions and Biased Masking",
  author        = "Han, Bin and Howe, Bill",
  month         =  jan,
  year          =  2023,
  archivePrefix = "arXiv",
  primaryClass  = "cs.CV",
  eprint        = "2301.04233"
}

@article{TemporalPaths,
 author = {Wu, Huanhuan and Cheng, James and Huang, Silu and Ke, Yiping and Lu, Yi and Xu, Yanyan},
 title = {Path Problems in Temporal Graphs},
% booktitle = {VLDB 2014},
 journal = {PVLDB},
 year = {2014},
 pages = {721--732},
}

@article{TACO_VLDB2022,
 author = {Fan, Wenfei and Jin, Ruochun and Lu, Ping and Tian, Chao and Xu, Ruiqi},
 title = {Towards Event Prediction in Temporal Graphs},
% booktitle = {VLDB 2022},
 journal = {PVLDB},
 year = {2022},
 pages = {1861–-1874},
}

@article{TemporalPathsJournal,
  author={H. {Wu} and J. {Cheng} and Y. {Ke} and S. {Huang} and Y. {Huang} and H. {Wu}},
  journal={TKDE},
  title={Efficient Algorithms for Temporal Path Computation},
  year={2016},
  volume={28},
  number={11},
  pages={2927-2942},
}

@INPROCEEDINGS{cilkplus,
  title     = "The Cilk++ concurrency platform",
  booktitle = "DAC",
  author    = "Leiserson, Charles E",
  pages     = "522--527",
  year      =  2009,
}

@INPROCEEDINGS{opencilk,
  title     = "{OpenCilk}: A Modular and Extensible Software Infrastructure for
               Fast {Task-Parallel} Code",
  booktitle = "PPoPP",
  author    = "Schardl, Tao B and Lee, I-Ting Angelina",
  pages     = "189--203",
  year      =  2023,
}

@misc{temporal_ligra,
  title = {Temporal Ligra},
  url = {https://github.com/jshun/ligra/tree/temporal}
}

@misc{Dataset_stackoverflow,
  title = {SNAP StackOverflow dataset},
  url = {https://snap.stanford.edu/data/sx-stackoverflow.html}
}

@inproceedings{ligra,
author = {Shun, Julian and Blelloch, Guy E. },
title = {Ligra: A Lightweight Graph Processing Framework for Shared Memory},
booktitle = {PPoPP},
year = {2013},
pages = {135--146}
}

@INPROCEEDINGS{packed_CSR,
  author={Wheatman, Brian and Xu, Helen},
  booktitle={HPEC}, 
  title={Packed Compressed Sparse Row: A Dynamic Graph Representation}, 
  year={2018},
  pages={1-7},
}

@inproceedings{TREND_WWW2022,
  author    = {Zhihao Wen and Yuan Fang},
  title     = {{TREND:} TempoRal Event and Node Dynamics for Graph Representation Learning},
%  booktitle = {{WWW} 2022},
  booktitle = {{WWW}},
  year      = {2022},
  pages     = {1159--1169},
}

@inproceedings{graphchi,
  title={GraphChi: Large-scale graph computation on just a PC},
  author={Kyrola, Aapo and Blelloch, Guy E and Guestrin, Carlos},
%  booktitle={OSDI 2012}
  booktitle={OSDI},
  year={2012},
  pages={31--46}
}


\end{document}